\documentstyle[prl,aps,twocolumn,epsfig,rotating]{revtex}
\begin{document}
\draft
\title{Observation of Magnetic Fingerprints in Superconducting
Au$_{0.7}$In$_{0.3}$ Cylinders}
\twocolumn[
\hsize\textwidth\columnwidth\hsize\csname@twocolumnfalse\endcsname
\author{Yu. Zadorozhny, D. R. Herman, and Y. Liu}
\address{Department of Physics, The Pennsylvania State University,
 University Park, PA 16802}
\date{\today}
\maketitle

\begin{abstract}
Reproducible, sample-specific magnetoresistance fluctuations (magnetic fingerprints) have been observed experimentally in the low-temperature part of the superconducting transition regime of disordered superconducting Au$_{0.7}$In$_{0.3}$ cylinders.  The amplitude of the fluctuation was found to exceed that of the universal conductance fluctuation in normal metals by several orders of magnitude.  The physical origin of these observations is discussed in the context of mesoscopic fluctuations of the superconducting condensation energy in disordered superconductors.
\end{abstract}

]
\narrowtext

In the past two decades fascinating phenomena in normal-metal mesoscopic systems have been found and, for the most part, understood \cite{1}.  One of the most important aspects of mesoscopic physics is quantum interference over a length much larger than the atomic size.  In disordered mesoscopic samples, this remarkable phenomenon is manifested in seemingly random but fully reproducible sample-specific magnetoresistance  fluctuations, referred to in literature as magnetic fingerprints (MFPs) \cite{1}.  These MFPs, which have emerged as a hallmark of mesoscopic physics, result from Aharonov-Bohm interference of electron waves.  Remarkably, the amplitude of the conductance fluctuations has a universal value of the order of $e^2/h$, known as the universal conductance fluctuation (UCF).  The physical origin of UCF lies in the energy level statistics in disordered metal, where the fluctuation in the number of energy levels within the Thouless energy is universally of the order of unity \cite{1}.

In the past few years, the UCF has also been examined in normal-metal samples in contact with one or more superconducting islands \cite{2}.  Andreev reflection from the normal metal-superconductor interfaces extends phase coherence in the normal metal beyond the normal coherence length \cite{3}, leading to new physical phenomena \cite{2}.  However, no significant change was found in conductance fluctuations \cite{4}, as anticipated theoretically \cite{5}.

Interesting questions arise if superconductivity is introduced in the bulk, rather than at the boundary of a normal sample.  Consider a weakly disordered mesoscopic sample in which electrons become phase coherent well above the onset of superconductivity.  These phase-coherent normal electrons are extremely sensitive to impurity scattering \cite{1}.  However, when electrons form Cooper pairs, they become completely insensitive to randomness.  How do electrons respond to these opposite tendencies of motion?  In addition, in disordered metallic samples, energy levels fluctuate, leading to MFPs and UCF as mentioned above.  What would the manifestation of the energy level fluctuation be in disordered superconductors?  Experimentally, these issues have not been examined prior to the present work.  In particular, no MFPs have been reported for superconductors.  In this Letter, we report results obtained on superconducting Au-In cylinders, in which sample-specific magnetoresistance fluctuations, or MFPs, have been found.

Superconducting Au-In binary alloy was originally chosen for this study because its critical temperature ($T_c$) can be easily controlled by changing the In concentration.  Au-In alloy has a rich phase diagram that includes compounds, AuIn and AuIn$_2$, and solid solutions with varying composition ratios \cite{alloys}.  For the latter, the $T_c$ continuously changes with In concentration \cite{Price}.  An important consequence of this is that inhomogeneity in In concentration results in spatially varying local $T_c$'s.  Spatial fluctuation in $T_c$ in our samples may be related to the origin of the observed sample-specific resistance fluctuations (see below).

\begin{figure}
\centerline{
\epsfig{file=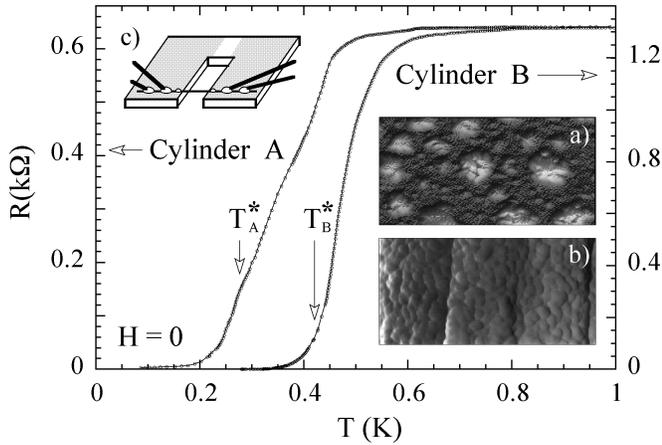,angle=0,width=3.4in}}
\medskip
\caption{Resistance as a function of temperature for Cylinders A and B in zero field.  Below $T = T^{\ast}$, the MR fluctuations were found.  Inset (a): an AFM image of a $4\times2$ $\mu$m$^2$ area of a 500\AA~thick planar Au$_{0.7}$In$_{0.3}$ film.  Inset (b): a flattened AFM image of a $1\times0.5$ $\mu$m$^2$ area of a cylindrical film.  The cylinder axis is vertical.  Inset (c): a schematic of the sample configuration.}
\label{1}
\end{figure}

In the {\it bulk} form, the maximum solid solubility of In in Au is about 12\% \cite{alloys}.  When the In concentration exceeds this limit, a phase separation is expected to occur, with the excess In forming In-rich grains.  In thick Au$_{0.7}$In$_{0.3}$ planar films these In-rich grains can be directly observed as they form micron-size grain conglomerates (Inset (a) of Fig. \ref{1}).  In thinner planar and cylindrical films such grain segregation is not found (Inset (b) of Fig. \ref{1}), probably because of the reduced mobility of atoms due to substrate effects.  Nonetheless, the onset of superconductivity, marked by the initial resistance decrease, occurs at the same temperature as in thicker films, suggesting that In-rich grains are also present in these samples \cite{films}.

To prepare a sample, an insulating (GE 7031 varnish) filament of submicron diameter was drawn and placed across a gap in a thin glass slide.  The slide was then mounted on a rotator inside an evaporation system.  A cylindrical film of Au$_{0.7}$In$_{0.3}$ was prepared by depositing 99.9999\% pure Au, In, and Au sequentially in the appropriate proportion onto the rotating filament.  The thickness of the films was measured with a quartz crystal thickness monitor.  The length of the free-standing cylindrical film is given by the width of the gap ($\approx$1mm).  The diameters of the cylinders were determined using scanning electron microscopy.  Current and voltage leads were attached to the cylinder using Ag epoxy.  The cylinders were manually aligned to be as parallel to the magnetic

\noindent field as possible.  The samples were stored at room temperature for at least several days and then slowly cooled down in a dilution refrigerator.  To make sure any possible residual thermal strain is relieved, the samples were kept at low temperature for several more days before any measurements were carried out.  {\it All} electrical leads entering the cryostat were RF filtered.  The resistance was measured in d.c. at 1$\mu$A.

In Fig. \ref{1}, resistance $R$ of two Au$_{0.7}$In$_{0.3}$ cylindrical films, Cylinders A and B, is plotted against temperature $T$.  The diameter, nominal thickness, and the normal-state sheet resistance of the two cylinders were respectively $0.84\mu$m, 350\AA, and $1.7\Omega$ for A and $0.60\mu$m, 300\AA, and $2.5\Omega$ for B.  A wide transition regime was found for both samples, as expected for inhomogeneous films.  The temperature range of the transition is consistent with the expected range of local $T_c$ variation, from $\approx 0.1$K for uniform Au$_{0.88}$In$_{0.12}$ matrix \cite{Price} to the maximum of $\approx 0.6$K ($T_c$ of bulk AuIn) for In-rich grains.  No resistance drop was seen at 3.4K, the superconducting transition temperature of bulk In.  The samples become fully superconducting around 0.1K for Cylinder A and 0.3K for Cylinder B.

\begin{figure}
\centerline{
\epsfig{file=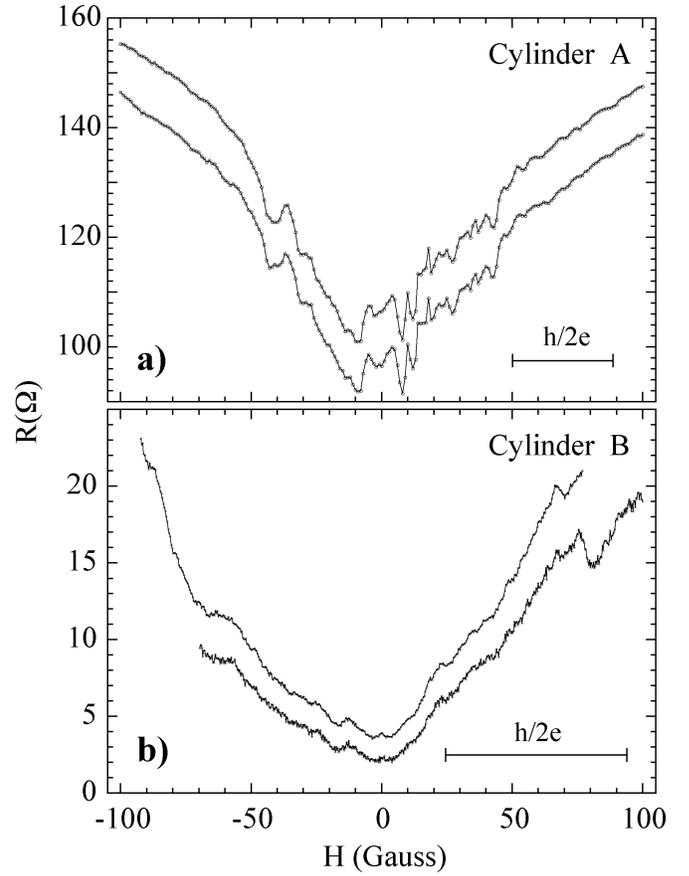,angle=0,width=3.4in}}
\medskip
\caption{a)  Two traces of MR scan for Cylinder A at $T = 0.25$K, upper trace offset by $10\Omega$.  b)  Two traces of MR scan for Cylinder B at $T = 0.35$K, upper trace offset by $1\Omega$.  Note that the bottom trace was measured at $0.5\mu$A.}
\label{2}
\end{figure}

In Fig. \ref{2}a, two traces of magnetoresistance (MR) scan for Cylinder A, taken at $T=0.25$K, deep into the superconducting transition regime, are shown.  A non-periodic, asymmetric (with respect to the reversal of the magnetic field) MR pattern is seen in both traces.  A comparison of the two traces shows a remarkable reproducibility of the pattern (the cross-correlation is 97\%).  This pattern can be seen as a reproducible resistance fluctuation, or magnetic fingerprint, in a positive, symmetric MR background expected for a superconductor.  Similar MFPs have been found in most Au$_{0.7}$In$_{0.3}$ cylinders we studied

\noindent so far.  In Fig. \ref{2}b we show a set of data obtained in Cylinder B.

\begin{figure}
\centerline{
\epsfig{file=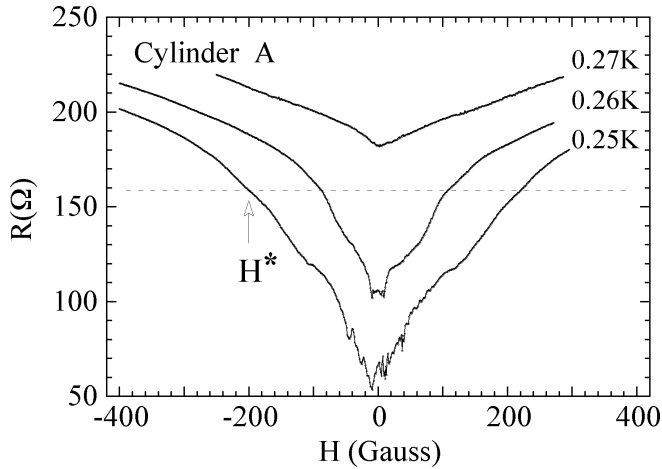,angle=0,width=3.4in}}
\medskip
\caption{MR traces of Cylinder A taken at temperatures as indicated.  MR fluctuations persist up to $H^{\ast}$.}
\label{3}
\end{figure}

A small increase in temperature was seen to suppress the fluctuations surprisingly strongly (Fig. \ref{3}).  For Cylinder A, at $T = T^{\ast} \approx 0.27$K, still deep in the transition regime, the resistance fluctuation already disappeared completely.  Magnetic field was found to have a similar effect.  Above a threshold field $H^{\ast}(T)$, the resistance fluctuation disappeared and the MR recovered the monotonic, symmetric behavior.  It is interesting to note that the fluctuation disappeared once the resistance was above certain value, either by increasing temperature or magnetic field.

\begin{figure}
\centerline{
\epsfig{file=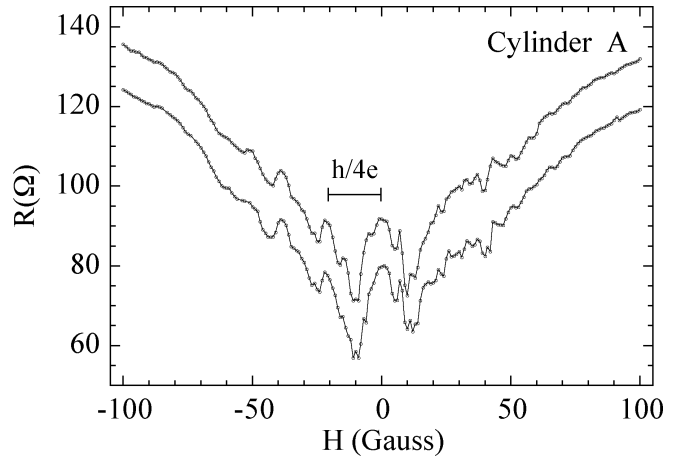,angle=0,width=3.4in}}
\medskip
\caption{Two MR traces for Cylinder A at $T = 0.25$K, featuring a different fluctuation pattern.  The upper trace is offset by $10\Omega$ for clarity.}
\label{4}
\end{figure}

The MFPs remained essentially the same in several consecutive scans.  However, after the sample was thermally cycled to around 10-15K, well above the onset of superconductivity, a different fluctuation pattern was found, as shown for example in Fig. \ref{3} (bottom trace) and in Fig. \ref{4} for Cylinder A.  The magnitude of the zero-field resistance $R_{H=0}$ at a fixed temperature was also found to change randomly as a result of thermal cycling.  The range of the resistance variation at $T=0.25$K was about 60$\Omega$, or 10\% of the normal-state resistance $R_N$.

\noindent Applying a high (several Tesla) magnetic field also irreversibly changed the MR, similar to thermal cycling.

A resistance maximum at zero magnetic field is clearly seen in Fig. \ref{4}.  Typically, superconducting fluctuations are suppressed by an  applied field, leading to a positive MR.  In the conventional Little-Parks (L-P) experiment, the MR at $H=0$ is always a minimum \cite{L-P}.  A negative MR as large as 25\% of $R_{H=0}$ deep in the superconducting transition regime is therefore very unusual.  Similar negative MR has been observed in other Au$_{0.7}$In$_{0.3}$ cylinders, as seen for example in Fig. \ref{2}b.  The negative MR was suppressed by a small temperature increase as all other features in the MR were (Fig. \ref{3}).

The data shown in Figs. \ref{2}-\ref{4} suggest that the conventional L-P resistance oscillation was too weak to be observed or even absent in Cylinders A and B, which we believe is due to the following reason.  All cylinders used in the present study were measured in a free-standing configuration.  Varnish undergoes a much larger thermal contraction than Au-In alloy.  Therefore "cracks" may have developed along the cylinder during the cooling down due to insufficient thermal anchoring of a free-standing sample, consistent with the AFM studies of some cylinders, which showed fine trenches parallel to the axis (Inset (b) of Fig. \ref{1}).  The multiply connected part of the cylinder may be small, with widely varying local $T_c$'s, leading to suppression of the L-P oscillation.

Sample-specific MR could in principle result from multiple magnetic field driven transitions if the sample consisted of a collection of superconducting weak links with varying local critical field.  In this picture, however, successive suppression of superconductivity of each individual weak link as the (parallel) field increases would result in monotonic, step-like features in MR, accompanied by hysteresis \cite{Adams}.  Instead, MR of our samples was found to be strongly non-monotonic and non-hysteretic.  Furthermore, the MR was asymmetric with respect to the magnetic field reversal, which also can not be explained

\noindent by the weak link picture.  All these considerations seem to suggest that superconducting weak links, if present in our samples, do not contribute significantly to the observed sample-specific MR.

Mesoscopic conductance fluctuations in {\it normal-metals} are sensitive to impurity configurations, magnetic fields, and gate voltages \cite{1}.  Thermal cycling to moderately high temperature can affect the impurity configuration and therefore result in a conductance change of the order of $e^2/h$.  Similarly, thermal cycling results in a conductance change in our samples.  It is possible that, due to uneven thermal contraction, thermal strains might have developed during temperature cycling.  Such strains could cause structural changes in the samples and might account for the observed variation in the sample resistance. However, thermal cycling did not affect the normal state resistance, suggesting that any resulted structural changes were very small.  Unlike in normal metals, conductance of our samples also changed irreversibly after they were exposed to a very high magnetic field, an issue that remains unresolved.  

In mesoscopic samples of normal metals, magnetic field also modifies the sample-specific conductance, resulting in MFPs.  Magnetic field of the order of the correlation field $H_{cor}$, corresponding to one flux quantum through the cross-section of the film, is required to change the conductance by $e^2/h$ \cite{1}.  MFPs were also found in our samples, however, due to the suppression of superconductivity, the MFPs were only observed in fields up to $H^{\ast}$, smaller than $H_{cor} \approx 450$G.  As a result the most prominent fluctuation features had field scale much smaller than $H_{cor}$.  It should be noted that conductance fluctuations on field scales much smaller than $H_{cor}$ have been observed in normal-metal samples \cite{6}, with amplitude smaller than $e^2/h$.

The similarities between the sample-specific conductance in our samples and in mesoscopic normal-metal systems strongly suggest that the observed features are mesoscopic in origin.  However, the amplitude of these sample-specific conductance fluctuations appears to be much larger than that observed in normal samples.  An order-of-magnitude estimate gives $\Delta G = \Delta R_\Box / R_\Box^2 \approx 10^4 \, e^2/h$ for Cylinder A at 0.25K, where $R_\Box$ is the sheet resistance of the sample.

Theoretically, significantly enhanced sample-specific conductance fluctuations have been predicted for {\it homogeneously} disordered superconductors in the transition regime.  It has been shown that under appropriate conditions, such as close to the superconductor-insulator transition or in a strong parallel magnetic field, fluctuations in superconducting condensation energy can be larger than its mean value \cite{Zhou}.  The physical origin of these exceedingly large fluctuations lies in the level statistics, precisely the origin of the UCF in normal metals.  The fluctuation in condensation energy will in turn manifest itself in fluctuations of the local $T_c$ even for a homogeneously disordered superconductor \cite{Z-B}.  Zhou and Biagini have shown that mesoscopic fluctuations of both Aslamasov-Larkin and Maki-Thompson contributions to conductivity would lead to a sample-specific conductance fluctuation above the $T_c$.  Because of the long-range phase coherence developing in superconductors as $T_c$ is approached, sample-specific conductance should be observable in arbitrarily large samples, as long as the temperature is sufficiently close to $T_c$.  Similar to normal samples, these fluctuations are sensitive to magnetic field, impurity configuration, and gate voltage.  Conductance fluctuations are greatly amplified due to the superconducting coherence resulted from Cooper pairing correlation, a spectacular example of quantum mesoscopic phenomena at a macroscopic scale.

The calculation of Zhou and Biagini has been carried out for homogeneously disordered superconductors.  Therefore, strictly speaking, it is not directly applicable to our experimental system.  Nonetheless, the salient features predicted by the theory are expected to be present for inhomogeneously disordered superconductors as well \cite{7}.  Below we compare our experimental observations with these predictions.  First, the predicted sample-specific conductance fluctuation was observed experimentally, and only in a narrow temperature range right above $T_c$, consistent with the theory.  Second, the amplitude of the sample-specific conductance fluctuation close to $T_c$ was found to greatly exceed that of the UCF in normal samples, again consistent with the theory.  Finally, negative MR, observed in our experiment, can be naturally accounted for in the same theoretical framework, as shown earlier by Spivak and Kivelson \cite{S-K}.  The qualitative agreement between our experimental observations and the theory strongly suggests that the same physics as discussed above is at work in our Au$_{0.7}$In$_{0.3}$ samples.

Before closing, we remark that the theory in reference \cite{S-K} also predicts a resistance oscillation of period $h/4e$, half of the L-P period.  This oscillation, if present in Figs. \ref{2}-\ref{3}, is masked by aperiodic features.  In Fig. \ref{4}, however, reproducible MR peaks appear around $H =$ 0, -20, and -40G.  The peak separation is very close to $h/4e$.  Therefore, these peaks could be a signature of an $h/4e$ resistance oscillation.  Finally, the observed asymmetry in the magnetoresistance may have a related physical origin.  In normal samples, only a 4-point measuring configuration would result in asymmetic MR \cite{8}, because of the fundamental requirement of time-reversal symmetry.  The Spivak-Kivelson theory \cite{S-K}, however, allows for time-reversal symmetry breaking in the ground state of a disordered superconductor, which may lead to asymmetric MR.

In conclusion, we have observed, for the first time, reproducible, sample-specific resistance fluctuations in disordered Au$_{0.7}$In$_{0.3}$ cylinders.  The amplitude of the fluctuation is much larger than that of the UCF in normal samples.  We have argued that the physical origin of these observations lies in the mesoscopic fluctuation of superconducting condensation energy, as predicted by theory.

The authors would like to acknowledge useful discussions with S. Kivelson, B. Spivak, and F. Zhou.  This work is supported by NSF under grant DMR-9702661.

\end{document}